\documentclass[allclo]{FBSart}

\usepackage{amsfonts}
\usepackage{amssymb}
\usepackage[dvips,final]{graphicx}

\usepackage{cite}                         
\usepackage{citesort}                     
\usepackage{multirow}                     
\usepackage{amsmath}
\usepackage{xspace}                       
\newcommand{\disc}{\discretionary{}{}{}}

\newcommand{\NPA}{\textnormal{Nucl.\ Phys.\ }\textbf{A}}

\newcommand{\half}{\frac{1}{2}}

\newcommand{\MeV}{\ensuremath{\mathrm{MeV}}}
\newcommand{\fm}{\ensuremath{\mathrm{fm}}}
\newcommand{\EFTNoPion}{EFT(${\pi\hskip-0.55em /}$)\xspace}

\newcommand{\NXLO}[1]{N\ensuremath{{}^{#1}}LO\xspace}

\newcommand{\wave}[3]{\ensuremath{{}^{#1}\mathrm{#2}_{#3}}}

\newcommand{\fourS}{\wave{4}{S}{\frac{3}{2}}}

\renewcommand{\Re}{\mathrm{Re}}
\renewcommand{\Im}{\mathrm{Im}}

\newcommand{\de}{\partial}

 \newcommand{\calK}{\mathcal{K}}

\title{How To Classify 3-Body Forces -- And Why}

\author{Harald W.~Grie\3hammer\thanks{\textit{E-mail address:}
    hgrie@ph.tum.de; present address: Universit\"at Erlangen.}  \thanks{9th
    November 2005. Preprint FAU-TP3-05/8, TUM-T39-05-14, nucl-th/0511039. To
    appear in Few-Body Systems.}}

\institute{Institut f\"ur Theoretische Physik III, Universit\"at
  Erlangen, D-91058 Erlangen, Germany
  \\\emph{and} 
  Institut f{\"u}r Theoretische Physik (T39), TU M{\"u}nchen, D-85747
  Garching, Germany}

\runningauthor{H.~W.~Grie\3hammer}

\runningtitle{How To Classify 3-Body Forces -- And Why}

\begin{document}
\maketitle
\begin{abstract}
  For systems with only short-range forces and shallow 2-body bound states,
  the typical strength of any 3-body force in all partial-waves, including
  external currents, is systematically estimated by renormalisation-group
  arguments in the Effective Field Theory of Point-Like Interactions. The
  underlying principle and some consequences in particular in Nuclear Physics
  are discussed. Details and a better bibliography in
  Ref.~\cite{Griesshammer:2005ga}.
\end{abstract}

\section{Introduction}

The Effective Field Theory (EFT) of Point-Like Interactions is a
model-independent approach to systems without infinite-range forces in Atomic,
Molecular and Nuclear Physics at very-low energies with shallow real or
virtual 2-body bound-states (``dimers''), see
e.g.~\cite{bedaque_bira_review,Braaten:2004rn} for reviews. When the size or
scattering length $a$ of a 2-body system is much larger than the size (or
interaction range) $R$ of the constituents, a small, dimension-less parameter
$Q=\frac{R}{a}$ allows to classify the typical size of neglected corrections
at $n$th order beyond leading order (\NXLO{n}) as about $Q^{n}$. For example,
$a\approx104$~\AA\ and $R\approx1$~\AA\ in the ${}^4\mathrm{He}_2$ molecule,
i.e.~$Q\approx\frac{1}{100}$, while $a\approx4.5\;\fm$ and $R\approx1.5$ in
the deuteron, i.e.~$Q\approx\frac{1}{3}$ in the ``pion-less'' EFT, \EFTNoPion,
where pion-exchange between nucleons is not resolved as non-local. Thus, the
detailed dynamics on the ``high-energy'' scale $R$ can vary largely: For
example, attractive van-der-Waals forces $\propto\frac{1}{r^6}$ balance in
${}^4\mathrm{He}_2$ a repulsive core generated by QED; but in Nuclear Physics,
one-pion exchange $\propto\frac{1}{r^{[1\dots3]}}$ is balanced by a
short-range repulsion whose origin in QCD is not yet understood. It is a
pivotal advantage of an EFT that it allows predictions of pre-determined
accuracy without such detailed understanding -- as long as one is interested
in low-energy processes, i.e.~Physics at the scale $a$, and not $R$. Even when
possible (as in QED -- at least at scales $\ge1\;\fm$), EFTs reduce
numerically often highly involved computations of short-distance contributions
to low-energy observables by encoding them into a few simple,
model-independent constants of contact interactions between the constituents.
These in turn can be determined by simpler simulations of the underlying
theory, or -- when they are as in QCD not (yet) tenable -- by fit to data.
Universal aspects of few-body systems with shallow bound-states are manifest
in EFTs, with deviations systematically calculable. In 2-body scattering for
example, the EFT of Point-Like Interactions reproduces the Effective-Range
Expansion, but goes beyond it in the systematic, gauge-invariant inclusion of
external currents, relativistic effects, etc.

Take 3-body forces (3BFs): They parameterise interactions on scales much
smaller than what can be resolved by 2-body interactions, i.e.~in which 3
particles sit on top of each other in a volume smaller than $R^3$.
Traditionally, they were often introduced \emph{a posteriori} to cure
discrepancies between experiment and theory, but such an approach is of course
untenable when data are scarce or 1- and 2-body properties should be extracted
from 3-body data.  But how important are 3BFs in observables? The
classification in EFTs rests on the tenet that a 3BF is included if and only
if necessary to cancel cut-off dependences in low-energy observables. I
outline this philosophy and its results in the following, finding that --
independent of the underlying mechanism -- 3BFs behave very much alike in such
disparate systems as molecular trimers and 3-nucleon systems, but do not
follow simplistic expectations.

\section{Construction}

\begin{figure}[!htb]
\begin{center}
  \includegraphics*[width=\textwidth]{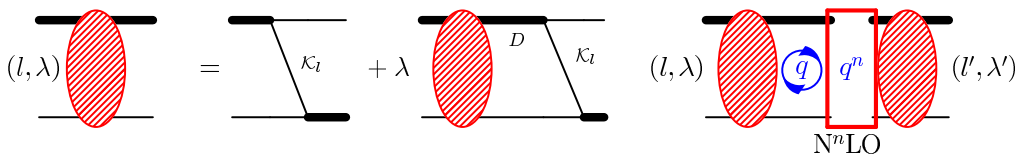}
  \caption{Left: integral equation of particle-dimer scattering. Right:
    generic loop correction (rectangle) at \NXLO{n}. Thick line ($D$):
    2-body propagator; thin line ($\calK_l$): propagator of the exchanged
    particle; ellipse: LO half off-shell amplitude $t_\lambda^{(l)}(p)$.}
\label{fig:kinematics}
\label{fig:higherorders}
\end{center}
\end{figure}

In the Faddeev equation of particle-dimer scattering without 3BFs,
Fig.~\ref{fig:kinematics}, the $\mathrm{S}$-wave 2-body scattering amplitude
is given by the LO-term of the Effective-Range Expansion.  This dimer and the
remaining particle ``interact'' via $\calK_l$, the one-particle propagator
projected onto relative angular momentum $l$.
Even for small relative on-shell momenta $k$ between dimer and particle, we
need the scattering amplitude $t_\lambda^{(l)}(p)$ for \emph{all} off-shell
momenta $p$ to determine its value at the on-shell point $p=k$, and hence in
particular for $p$ beyond the scale $\frac{1}{R}$ on which a description in
terms of point-like constituents is tenable. It is therefore natural to demand
that all low-energy observables on a scale $k\sim\frac{1}{a}$ are insensitive
to derails of the amplitude at $p\gg\frac{1}{R}$, namely to form and value of
the regulator, form-factor or cut-off chosen.  If not, a 3BF must soak up the
dependence.  In an EFT, this is the fundamental tenet: Include a 3BF \emph{if
  and only if} it is needed as counter-term to cancel divergences which can
not be absorbed by renormalising 2-body interactions. Thus, only combinations
of 2- and 3BFs are physically meaningful. With the cut-off variation of the
3BF thus fixed, the initial condition leads to one free parameter fixed from a
3-body datum or knowledge of the underlying physics.  3BFs are thus not added
out of phenomenological needs but to guarantee that observables are
insensitive to off-shell effects.

A Mellin transformation $t_\lambda^{(l)}(p)\propto p^{-s_l(\lambda)-1}$ solves
the equation for $p\gg k,\frac{1}{a}$.  The spin-content is then encoded only
in the homogeneous term: $\lambda=-\half$ for 3 nucleons with total spin
$\frac{3}{2}$, or for the totally spin and iso-spin
\mbox{(Wigner-)}\disc{}anti-symmetric part of the spin-$\frac{1}{2}$-channel;
$\lambda=1$ for 3 identical spin-less bosons and the totally spin and iso-spin
\mbox{(Wigner-)}\disc{}symmetric part of the spin-$\frac{1}{2}$-channel.  The
asymptotic exponent $s_l(\lambda)$ has to fulfil $\Re[s]>-1$,
$\Re[s]\not=\Re[l\pm2]$, and
\begin{equation}
  \label{eq:s}
  1=\left(-1\right)^l\;\frac{2^{1-l}\lambda}{\sqrt{3\pi}}\;
  \frac{\Gamma\left[\frac{l+s+1}{2}\right]\Gamma\left[\frac{l-s+1}{2}\right]}
  {\Gamma\left[\frac{2l+3}{2}\right]}\;
  {}_2F_1\left[\frac{l+s+1}{2},\frac{l-s+1}{2};
    \frac{2l+3}{2};\frac{1}{4}\right].
\end{equation}   
This result was first derived in the hyper-spherical approach by Gasaneo and
Macek~\cite{lost}\footnote{My apologies to the authors that I found this
  reference only after \cite{Griesshammer:2005ga} was published.}.  The
asymptotics depends thus only but crucially on $\lambda$ and $l$. Relevant in
the UV-limit are the solutions for which $\Re[s+1]$ is minimal.

\begin{figure}[!ht]
\begin{center}
  \includegraphics*[width=0.48\textwidth]{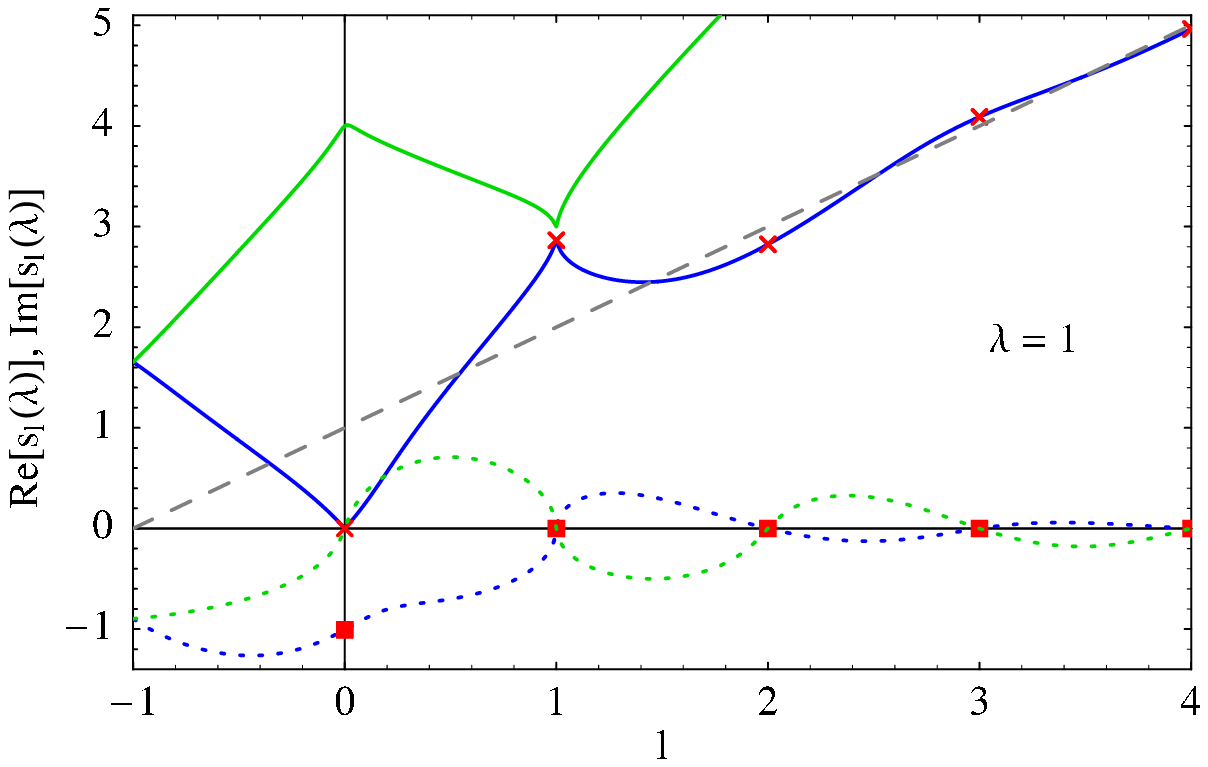}\hfill
  \includegraphics*[width=0.48\textwidth]{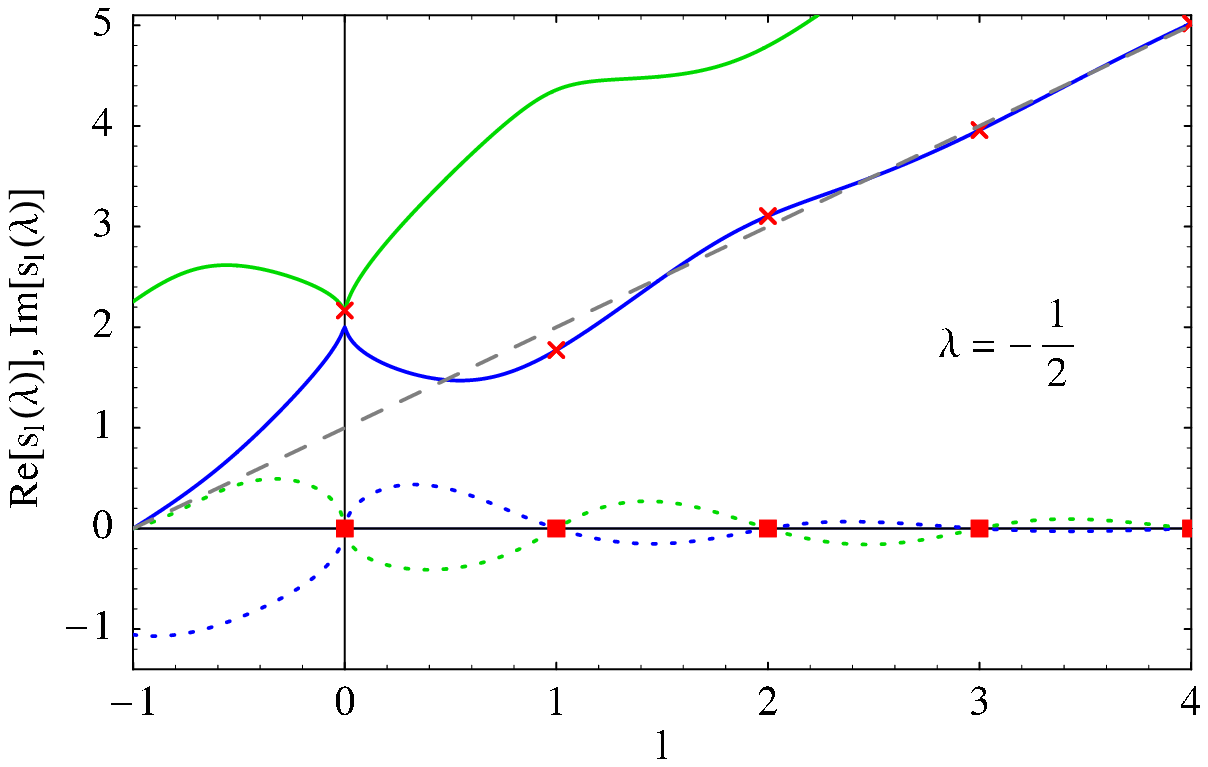}
\caption{The first two solutions $s_l(\lambda)$ at $\lambda=1$ (left) and
  $\lambda=-\half$. Solid (dotted): real (imaginary) part; dashed: simplistic
  estimate. 
  Dark/light: first/second solution. Limit cycle and Efimov effect occur only
  when the solid line lies below the dashed one, and $\Im[s]\not=0$.}
\label{fig:s-lambdafixed}
\end{center}
\end{figure}

At first glance, we would expect the asymptotics to be given by the
asymptotics of the inhomogeneous (driving) term:
$t_\lambda^{(l)}(p)\stackrel{?}{\propto}\frac{k^l}{p^{l+2}}$,
i.e.~$s_{l}(\lambda)\stackrel{?}{=}l+1$. However, we must sum an infinite
number of graphs already at leading order. As Fig.~\ref{fig:s-lambdafixed}
shows, this modifies the asymptotics considerably.

How sensitive are higher-order corrections to the UV-behaviour of
$t_\lambda^{(l)}(p)$?  The 2-body scattering-amplitude is systematically
improved by including the effective range, higher partial waves etc.
Corrections to 3-body observables (including partial-wave mixing) are found by
perturbing around the LO solution as in Fig.~\ref{fig:kinematics}. Most
sensitive to unphysically high momenta is each correction at \NXLO{n} which is
proportional to the $n$th power of loop momenta. The question when it becomes
cut-off sensitive is now rephrased as: When does the correction diverge as the
cut-off is removed, i.e.~when is its \emph{superficial degree of divergence}
non-negative? The answer by simply counting loop momenta in the diagram:
\begin{equation}
  \label{eq:twobodydivs}
  \Re[n-s_l(\lambda)-s_{l^\prime}(\lambda^\prime)]\geq0\;\;.
\end{equation}
We therefore find at which order the first 3BF is needed just by determining
when a correction to the 3-body amplitude with only 2-body interactions
becomes dependent on unphysical short-distance behaviour.

It is instructive to re-visit these findings in position space.  The
Schr\"odinger equation for the wave-function in the hyper-radial
dimer-particle distance $r$,
\begin{equation}
  \label{eq:hyperradial}
  \left[-\frac{1}{r}\;\frac{\de}{\de r}\;r\;\frac{\de}{\de r}+
  \frac{s_l^2(\lambda)}{r^2}-ME\right] F(r)=0\;\;,
\end{equation}
looks like the one for a free particle with centrifugal barrier. One would
thus expect $s_l\stackrel{?}{=}l+1$ (hyper-spherical co-ordinates!). It had
however already been recognised by Minlos and Faddeev that the centrifugal
term is for three bosons ($\lambda=1$) despite expectations attractive, so
that the wave-function collapses to the origin and seems infinitely sensitive
to very-short-distance physics. In order to stabilise the system against
collapse -- or, equivalently, remove dependence on details of the cut-off --,
a 3BF must be added -- or, equivalently, a self-adjoint extension be specified
at the origin i.e.~a boundary condition for the wave-function must be fixed by
a 3-body datum. On the other hand, 3BFs are demoted if $s_l> l+1$: The
centrifugal barrier provides \emph{more} repulsion than expected, and hence
the wave-function is pushed further out, i.e.~\emph{less} sensitive to details
at distances $r\lesssim R$ where the constituents are resolved as extended.
Birse confirmed this recently by a renormalisation-group analysis in
position-space~\cite{Birse:2005pm}.

\section{Consequences}

About half of the 3BFs for $l\leq 2$ are \emph{weaker}, half \emph{stronger}
than one would expect simplistically, see Table~\ref{tab:ordering}. The higher
partial-waves follow expectation, as the Faddeev equation is then saturated by
the Born approximation.
\begin{table}[!ht]
{\small
\begin{tabular}{|cc||c|c||c||c@{\hspace*{3pt}}c|}
  \hline
\multicolumn{2}{|c||}{channel}&  na\"ive dim.~analysis &simplistic&&
\multicolumn{2}{|c|}{typ.~size}\\
  bosons& fermions&{$\mathrm{Re}[s_l(\lambda)+s_{l^\prime}(\lambda^\prime)]$} &
  {$l+l^\prime+2$}&&\multicolumn{2}{|c|}{if $Q^n\sim\frac{1}{3^n}$}\\
  \hline\hline
  \wave{}{S}{}-\wave{}{S}{}&\wave{2}{S}{}-\wave{2}{S}{}&{LO}&
\NXLO{2}&{prom.}&{$100\%$}& ($10\%$)\\
  &\wave{2}{S}{}-\wave{4}{D}{}&\NXLO{3.1}&\NXLO{4}&prom.&{$3\%$}& ($1\%$)\\
\hline\hline
\wave{}{P}{}-\wave{}{P}{}&\wave{2}{P}{}-\wave{2}{P}{}&\NXLO{5.7}&&dem.&0.2\%&\\
&\hspace*{-1ex}\wave{2}{P}{}-\wave{2}{P}{},
\wave{4}{P}{}-\wave{4}{P}{}&\NXLO{3.5}&\NXLO{4}&prom.&{$2\%$} &($1\%$)\\
&\wave{2}{P}{}-\wave{4}{P}{}  &\NXLO{4.6}&& dem.&{$0.6\%$}& \\
\hline\hline
&\wave{4}{S}{}-\wave{4}{S}{}&{\NXLO{4.3+2}}&{\NXLO{2+2}}&{dem.}&{$0.1\%$}& \\
&\wave{4}{S}{}-\wave{2}{D}{}&\NXLO{5.0}&
  \multirow{2}{2cm}{\hspace*{\fill}\NXLO{4}\hspace*{\fill}}&dem.&{$0.4\%$}&
  ($1\%$)\\ 
&\wave{4}{S}{}-\wave{4}{D}{}&\NXLO{5.3}&&dem.&{$0.3\%$}&\\
\hline
\multicolumn{2}{|c||}{higher}&{$\sim$ as simplistic}&{\NXLO{l+l^\prime+2}}&&&\\
\hline
  \end{tabular}
}
\caption{Order of the leading 3BF, indicating if actual
  values (\ref{eq:twobodydivs}/\ref{eq:s}) are stronger (``prom.'') or
  weaker (``dem.'') than the simplistic  estimate. 
  Last column: typical size of 3BF in \EFTNoPion; in parentheses size from the
  simplistic estimate.}
  \label{tab:ordering}
\end{table}
The $\mathrm{S}$-wave 3BF of spin-less bosons is stronger, while the
$\mathrm{P}$-wave 3BF is weaker.

That the first $\mathrm{S}$-wave 3BF appears already at LO leads to a new
renormalisation-group phenomenon, the ``limit-cycle''. It explains the Efimov
and Thomas effects, and universal correlations e.g.~between particle-dimer
scattering length and trimer binding energy (the Phillips line).  In general,
it appears whenever the kernel of the integral equation not compact,
i.e.~$\Im[s]\not=0$ and $|\Re[s]|<\Re[l+1]$.
We finally note that the power-counting requires a new, independent 3BF with
$2l$ derivatives to enter at \NXLO{2l} and provides high-accuracy phase-shifts
in atom-dimer and nucleon-deuteron scattering, and loss rates close to
Feshbach resonances in Bose-Einstein condensates, see
e.g.~\cite{improve3body,bedaque_bira_review,Braaten:2004rn} for details.

Demotion might seem an academic dis-advantage -- to include some higher-order
corrections which are not accompanied by new divergences does not improve the
accuracy of the calculation; one only appears to have worked harder than
necessary. However, demotion is pivotal when one wants to predict the
experimental precision necessary to dis-entangle 3BFs in observables, and here
the error-estimate of EFTs is crucial.
In ${}^4\mathrm{He}$-atom-dimer scattering, where $Q\approx\frac{1}{100}$,
only high-precision experiments can however reveal contributions from 3BFs
beyond the one found in the $\mathrm{S}$-wave.

On the other hand, $Q\approx\frac{1}{3}$ in \EFTNoPion of Nuclear Physics.
Now, the demotion or promotion of 3BFs makes all the difference whether an
experiment to determine 3BF effects is feasible at all.  For example, the
quartet-$\mathrm{S}$ wave scattering-length in the neutron-deuteron system
sets at present the experimental uncertainty in an indirect determination of
the doublet scattering length, which in turn is well-known to be sensitive to
3BFs.  Its value in \EFTNoPion at \NXLO{2},
 \begin{equation}
   a(\fourS)=[
   5.09(\mathrm{LO})+1.32(\mathrm{NLO})-0.06(\mathrm{\NXLO{2}})]\;\fm=
   [6.35\pm0.02]\;\fm\;\;,
 \end{equation} 
 converges nicely and agrees very well with experiment, $[6.35\pm0.02]\;\fm$.
 The theoretical accuracy by neglecting higher-order terms is here estimated
conservatively by $Q\approx\frac{1}{3}
$ of the difference between the NLO- and \NXLO{2}-result.
Table~\ref{tab:ordering} predicts that the first 3BF enters not earlier than
\NXLO{6}. Indeed, if the theoretical uncertainty continues to decreases
steadily as from NLO to \NXLO{2}, an accuracy of
$\pm(\frac{1}{3})^3\times0.02\;\fm\lesssim\pm0.001\;\fm$ with only 2-nucleon
scattering data as input can be reached in calculations.
This is comparable to the range over which modern high-precision
potential-model calculations differ: $[6.344\dots6.347]\;\fm$.  If the 3BF
would occur at \NXLO{4} as simplistically expected, the error by 3BFs would be
$(\frac{1}{3})^1\times0.02\;\fm\approx0.007\;\fm$, considerably larger than
the spread in the potential-model predictions.  Differential cross-sections
and partial-waves are also in excellent agreement with much more elaborate
state-of-the-art potential model calculations at energies up to $15\;\MeV$,
see e.g.~\cite{improve3body}.

The cross-section of triton radiative capture $nd \to t\gamma$ at thermal
energies provides another example~\cite{withSadeghi}. Nuclear Models give a
spread of $[0.49\dots0.66]\;\mathrm{mb}$, depending on the 2-nucleon
potential, and how the $\Delta(1232)$ as first nucleonic excitation is
included. On the other hand, a process at $0.0253$ eV [\emph{sic}] incident
neutron energy and less than $7\;\MeV$ photon energy should be insensitive to
details of the deuteron wave-function and of a resonance with an excitation
energy of $300\;\MeV$. Indeed, the power-counting of 3BFs applies equally with
external currents, only that the higher-order interaction in
Fig.~\ref{fig:higherorders} includes now also the momentum- or energy-transfer
from the external source as additional low-energy scales. As no new 3BFs are
needed up to \NXLO{2} to render cut-off independent results, the result is
completely determined by simple 2-body observables as
\begin{equation}
   \sigma_\mathrm{tot}=[
   0.485(\mathrm{LO})+0.011(\mathrm{NLO})+0.007(\text{\NXLO{2}})
   ]\;\mathrm{mb}=
   [0.503\pm0.003]\;\mathrm{mb}\;\;,
\end{equation}
which converges and compares well with the measured value,
$[0.509\pm0.015]\;\mathrm{mb}$. 
The cross-section relevant for big-bang nucleo-synthesis
($E_n\approx0.020\dots0.4\;\mathrm{MeV}$) is also in excellent agreement with
data~\cite{Sadeghi:2004es}.

 
\section{Conclusions}

With these findings, the EFT of 3-body systems with only contact interactions
is a self-consistent, systematic field theory which contains the minimal
number of interactions at each order to render the theory renormalisable. Each
3-body counter-term gives rise to one subtraction-constant which is fixed by a
3-body datum.  Table~\ref{tab:ordering} sorts the 3BFs by their strengthes,
their symmetries and the channels in which they contribute at the necessary
level of accuracy.  Amongst the host of applications in Nuclear Physics are
triton and ${}^3$He properties, reactions in big-bang nucleo-synthesis,
neutrino astro-physics, the famed nuclear $A_y$-problem, and the experimental
determination of fundamental neutron properties.

The method presented here is applicable to any EFT in which an infinite number
of diagram needs to be summed at LO, e.g.~because of shallow bound-states. One
example is Chiral EFT, the EFT of pion-nucleon interactions. Only those local
$N$-body forces are added at each order which are necessary as counter-terms
to cancel divergences at short distances.  This mandates a careful look at the
ultraviolet-behaviour of the leading-order, non-perturbative scattering
amplitude. It leads at each order and to the prescribed level of accuracy to
a cut-off independent theory with the smallest number of experimental
input-parameters. The power-counting is thus not constructed by educated
guesswork but by rigorous investigations of the renormalisation-group
properties of couplings and observables using the methodology of EFT.

\begin{acknowledge}
My thanks to the organisers for creating a highly stimulating workshop and the
warm welcome. Supported in part by DFG Sachbeihilfe GR 1887/2-2, 3-1, and BMFT.
\end{acknowledge}

\end{document}